# Non-intrusive Load Monitoring based on Self-supervised Learning

Shuyi Chen, *Student Member*, *IEEE*, Bochao Zhao, *Member*, *IEEE*, Mingjun Zhong, *Member*, *IEEE*, Wenpeng Luan\*, *Senior Member*, *IEEE*, and Yixin Yu, *Life Senior Member, IEEE*

*Abstract*—Deep learning models for non-intrusive load monitoring (NILM) tend to require a large amount of labeled data for training. However, it is difficult to generalize the trained models to unseen sites due to different load characteristics and operating patterns of appliances between data sets. For addressing such problems, self-supervised learning (SSL) is proposed in this paper, where labeled appliance-level data from the target data set or house is not required. Initially, only the aggregate power readings from target data set are required to pre-train a general network via a self-supervised pretext task to map aggregate power sequences to derived representatives. Then, supervised downstream tasks are carried out for each appliance category to fine-tune the pre-trained network, where the features learned in the pretext task are transferred. Utilizing labeled source data sets enables the downstream tasks to learn how each load is disaggregated, by mapping the aggregate to labels. Finally, the fine-tuned network is applied to load disaggregation for the target sites. For validation, multiple experimental cases are designed based on three publicly accessible REDD, UK-DALE, and REFIT data sets. Besides, state-of-the-art neural networks are employed to perform NILM task in the experiments. Based on the NILM results in various cases, SSL generally outperforms zero-shot learning in improving load disaggregation performance without any sub-metering data from the target data sets.

## I. Introduction

IN recent years, energy shortage and environmental pollution worldwide have become increasingly serious. Therefore, the approaches of efficient energy utilization and carbon emissions reduction are being explored [1], [2]. Meanwhile, with the global deployment of smart meters, benign interaction between power suppliers and users has been established for enhancing demand side management and optimizing power grid operation [3]. As one of the energy conservation applications, electricity consumption detail monitoring has attracted extensive attention around the world [4]. In general, load monitoring technology is mainly categorized into intrusive way and non-intrusive way. Note that intrusive load monitoring requires extra sensor installation for sub-metering. Alternatively, the concept of non-intrusive load monitoring (NILM) was proposed by Hart [5] in 1984 as identifying power consumed by each individual appliance via analyzing aggregate power readings using only software tools. NILM offers appliance-level power consumption feedback to both demand and supply sides economically and efficiently, contributing to power system planning and operation [1], energy bill savings [6], demand side management [7], energy conservation and emission reduction [3], [6], [8], etc.

NILM is a single-channel blind source separation problem, aiming to disaggregate the appliance-level energy consumption from the aggregate measurements [9]. Combinatorial optimization (CO) is initially applied to perform NILM in [5], searching for the best combination of operational states of individual appliances at each time instance. However, CO relies on the power range of each operational state as prior knowledge, making it unavailable to the newly added appliances [10]. Benefiting from the technology development in recent years on big data, artificial intelligence and edge computing, plenty of NILM approaches have been proposed based on machine learning, mathematics, and signal processing [8], [11]. Factorial hidden Markov model (FHMM) and its variants [12]-[14] are popular in carrying out NILM. Given an aggregate power signal as the observation, such FHMM-based NILM methods estimate the hidden operational states of each appliance considering their state continuity in time-series [15], [16]. Thus, FHMM-based methods usually achieve good results in disaggregating loads with periodic operation such as refrigerators. However, their performance is limited for the loads with short-lasting working cycles and the ones with less frequent usage. Note that FHMM-based methods are regarded as state-based NILM approaches, where the aggregate power measurement at each time instance is assigned to each operational state per appliance [17]. Alternatively, NILM approaches can be event-based, where sudden changes in power signals referring to turn-on, turn-off, and state transition events are featured [17]. Such event-based NILM methods can be carried out via subtractive clustering and the maximum likelihood classifier [18]. Besides, graph signal processing concepts are applied to perform NILM, mapping correlation among samples to the underlying graph structure [19], [20]. Although such event-based NILM approaches can achieve high load identification accuracy, they tend to suffer from

This work was supported in part by the Joint Funds of the National Natural Science Foundation of China (No. U2066207) and the National Key Research and Development Program of China (No. 2020YFB0905904). *(Corresponding author: W. Luan)*

S. Chen, B. Zhao, W. Luan and Y. Yu is with the School of Electrical and Information Engineering, Tianjin University, Tianjin 300072, China (e-mail: wenpeng.luan@tju.edu.cn).

M. Zhong is with the Department of Computing Science, University of Aberdeen, Aberdeen, the UK (e-mail: mingjun.zhong@abdn.ac.uk).

measurement noises.

Deep neural networks (DNN), performing well in computer vision, speech recognition, and natural language processing, have been employed in load disaggregation since 2015 [21]. Since then, DNN-based NILM approaches become more and more popular, including long short-term memory (LSTM) [15], [21], gated recurrent unit (GRU) [10], [22], denoising autoencoder (dAE) [21], [23] and convolutional neural network (CNN) [24], [25], etc., showing competitive performance against traditional NILM methods. Although LSTM is suitable for long time-series related tasks due to avoiding the vanishing gradient problem, it underperforms in NILM task compared to CNN [26], [27]. As a variant of LSTM, GRU can also remember data patterns. In addition, GRU contains fewer parameters, thus requires shorter training time, which is suitable for online application in NILM [10], [28]. Note that the bidirectional gated recurrent unit (Bi-GRU) is employed to perform NILM in [10], where the network can be trained simultaneously in positive and negative time directions. Besides, dAE is applied to NILM by recovering the power signal of target appliances (clean signal) from the aggregate (noisy signal) [8], [21], where CNN layers are usually embedded [21], [23]. A state-of-the-art CNN-based NILM method, S2p, is proposed in [9] and claimed to outperform benchmarks in NILM task [9], [29], beneficial from meaningful latent features learned from sub-metering data. Compared to traditional NILM methods, advantages of DNN-based NILM approaches include automatic feature extraction from power readings and linearity between computational complexity and appliance amount [4]. However, the promising performance of the aforementioned DNN-based methods relies on a large amount of sub-metering data from the target set for training [4]. Since such data collection may last for months or even years [4], it is neither user-friendly nor economical in practice.

Alternatively, transfer learning concepts are proposed, where transferable networks can be trained on a source (seen) data set and applied to the load disaggregation task on a target (unseen) data set [2]. Depending on whether network fine-tuning is required, transfer learning can be classified as few-shot learning (FSL) and zero-shot learning (ZSL) [30]. For fine-tuning in FSL, a small amount of labeled data from the target set is still required [30]. However, when labels are unable to be captured from the target data sets, ZSL offers proper solutions. In [31], ZSL achieves tiny performance drop compared to baseline when it is employed in load disaggregation by both GRU and CNN networks, showing transferability across data sets. However, in ZSL, it is difficult to generalize networks between data sets with different load characteristics and operating patterns of appliances. The same as ZSL, self-supervised learning (SSL) requires no labeled data from the target set. SSL is an efficient way to extract universal features from large-scale unlabeled data, contributing to robustness enhancement [32], thus it performs well in image processing and speech recognition [33]. To the best of our knowledge, SSL has not been used to solve NILM problem.

Driven by such research gaps, in this paper, SSL is applied to two state-of-the-art NILM algorithms based on CNN and GRU, as S2p [9] and Bi-GRU [10]. For performing NILM, a self-supervised pretext task training is initially carried out for learning features from the aggregate power readings from the unlabeled data in the target set. Then the pre-trained network is fine-tuned in the supervised downstream task training based on the labeled data from the source set for transferring the pre-learned knowledge to load disaggregation. After pre-training and fine-tuning, the network can be applied to load disaggregation for target sites. The proposed method is validated on the real-world data sets at 1-min granularity, in the scenarios designed for the same data set or across various data sets. The contributions of this paper are clarified as follows:

- SSL is applied to load disaggregation based on deep learning without sub-metering on the target set, by setting a pretext task for network pre-training on unlabeled data from the target set with fine-tuning.
- Experiments are carried out for all combinations of two state-of-the-art DNN-based NILM methods (S2p [9] and Bi-GRU [10]) and learning frameworks (SSL with various fine-tuning ways and ZSL), on three real-world data sets;
- Six cases differing in data selection are designed for performance evaluation on the data across houses or sets, showing SSL generally outperforms in various metrics and energy consumption estimation results, with comparable training time cost.

The rest of this paper is organized as follows: in Section II, the NILM formulation is clarified, followed by introducing the preliminaries for NILM neural networks and SSL; The methodology of SSL for NILM is explained in Section III; Section IV contains data sets, evaluation metrics, and experimental settings, followed by experimental results with discussion illustrated in Section V; eventually, the conclusion is drawn and the future work is prospected in Section VI.

## II. PRELIMINARIES

In this section, we first formulate the NILM problem and then clarify seq2seq and seq2point concepts, followed by an introduction for two seq2point network architectures. Finally, the overall structure of SSL is demonstrated.

*A. NILM Problem Formulation*

Assuming that the aggregate power reading measured in a household at time index $t \in [1, T]$ is $y_t$, where $T$ refers to the total number of samples. Then the simultaneous power consumed by appliance $m \in \mathcal{M}$ to be disaggregated is denoted by $x_t^m$. The measurement noise is denoted by $e_t$, usually regarded as Gaussian distributed [7]. Then the total load power for a household can be expressed as:

$$y_t = \sum_{m=1}^{M} x_t^m + e_t. \qquad (1)$$

Thus, for each time index $t$, NILM problem is to estimate $x_t^m$, $\forall m$, given the aggregate power $y_t$. When applying machine learning or deep learning to NILM, it will become a regression or classification problem [7].

*B. Sequence-to-sequence vs. Sequence-to-point NILM Frameworks*

NILM can be carried out via neural networks with a seq2seq or seq2point framework [24]. In a seq2seq NILM solution, for each appliance, a network learns the non-linear regression between sequences with the same time stamps, referring to the aggregate and appliance-level power. For an arbitrary aggregate power sequence $\mathbf{y}$ covering time instance $t$, the power $x_t^m$ consumed by appliance $m$ at $t$ is predicted by the network, thus can be finalized as the average value of all such predictions [9]. Unlike the seq2seq framework, the seq2point framework predicts the appliance-level power consumed at only one point of each sliding window iteratively. The inputs and outputs of both seq2seq and seq2point frameworks in a NILM task are illustrated in Fig.1.

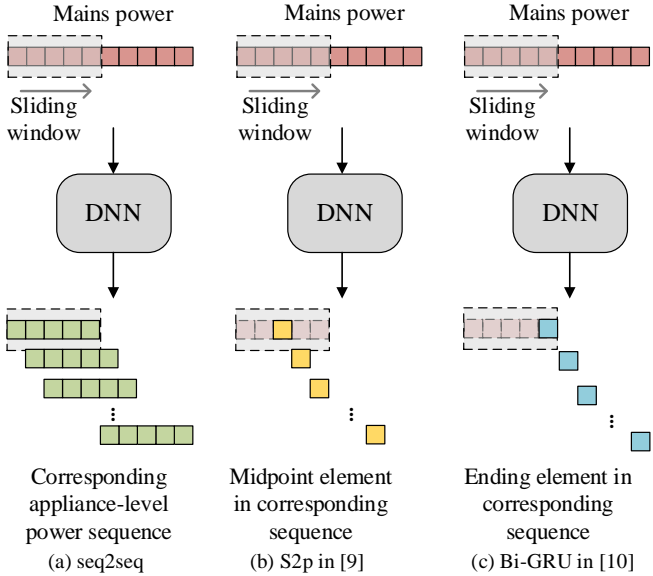

**Fig. 1.** Examples for seq2seq and seq2point frameworks.

Note that seq2point frameworks are demonstrated in Fig. 1 (b) and (c) on two architectures, as S2p proposed in [9] and Bi-GRU proposed in [10], respectively. Compared to the seq2seq framework, the seq2point framework emphasizes the representational power at one element and eases the prediction task. Then, S2p and Bi-GRU are introduced in details.

*1) S2p:* The utilization of S2p in NILM is based on the assumption that the midpoint of each sliding window acts as its non-linear regression representation. Namely, S2p makes full use of the past and future information to infer the midpoint, as shown in Fig. 1 (b).

For a defined neural network $f^m$, the input is a power sequence denoted by $\mathbf{y}_{t:t+W-1}$ segmented by a sliding window from the aggregate, where $t$ is time index and the window size $W$ is set to an odd number. Thus, by mapping each sequence $\mathbf{y}_{t:t+W-1}$ to the power $x_\tau^m$ consumed by appliance $m$ at $\tau = t + (W-1)/2$, the entire power signal $\mathbf{x}^m$ for appliance $m$ can be predicted. Such model can be formulated as:

$$x_\tau^m = f^m(\mathbf{y}_{t:t+W-1}) + \varepsilon, \qquad (2)$$

where $\varepsilon$ is $W$-dimensional Gaussian random noise. Besides, the loss function in the network training is formulated:

$$L_p = \sum_{t=1}^{T-W+1} \log p(x_\tau^m \mid \mathbf{y}_{t:t+W-1}, \theta_p), \qquad (3)$$

where $\theta_p$ is a set of network parameters.

The CNN-based architecture of S2p is illustrated in Fig. 2 (a) containing five convolutional layers and one dense layer. In each iteration, the input signal refers to an $n$-length sliding window for aggregate measurements. Then, five convolutional layers are employed for feature extraction through an activation function called ReLU. Eventually, the feature maps are flattened and fed to a dense layer, and an appliance-level power corresponding to the midpoint of the input window is obtained. It is claimed in [29] that S2p achieves performance improvement against seq2seq framework on the same network architecture.

*2) Bi-GRU:* Unlike S2p in Fig. 1 (b), the power consumed by each appliance at time index $t + W - 1$ is mapped by the pre-defined sequence $\mathbf{y}_{t:t+W-1}$ in Bi-GRU as historical aggregate measurements, shown in Fig. 1 (c). That is, as window sliding, power prediction per appliance can be obtained from only the past information, which is applicable for real-time load disaggregation. Moreover, GRU is beneficial from less memory occupancy and fewer parameters than other network architectures such as LSTM [28]. The architecture of Bi-GRU is demonstrated in Fig. 2 (b).

As shown in Fig. 2 (b), after each aggregate power sequence is input to the network, a convolutional layer is used for feature extraction. Then two Bi-GRU layers are applied to enhance the memory for the data patterns based on the extracted features, followed by a dense layer as in the S2p network. Note that dropout performs overfitting prevention for such layers. The network specifications for both S2p and Bi-GRU are clarified in Table I and Table II, respectively.

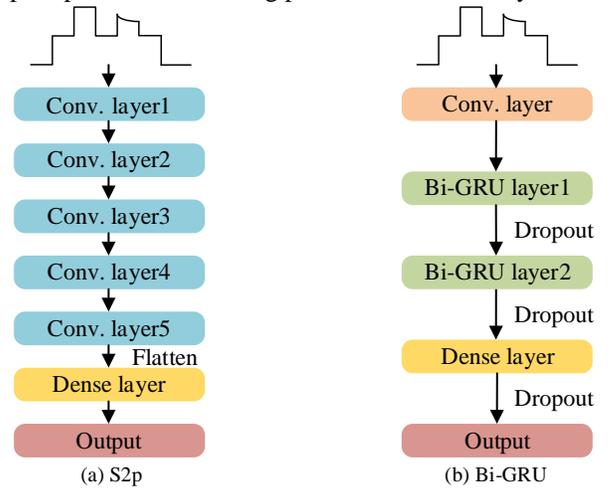

**Fig. 2.** The architectures for S2p and Bi-GRU.

TABLE I
NETWORK SPECIFICATIONS FOR S2P [9]

| Layer | Network specifications | | | |
|---|---|---|---|---|
| Input | $N$-length sequence | | | |
| Conv. layer1 | Filter size: 10 | # filters: 30 | Stride: 1 | Activation: ReLU |
| Conv. layer2 | Filter size: 8 | # filters: 30 | Stride: 1 | Activation: ReLU |
| Conv. layer3 | Filter size: 6 | # filters: 40 | Stride: 1 | Activation: ReLU |
| Conv. layer4 | Filter size: 5 | # filters: 50 | Stride: 1 | Activation: ReLU |
| Conv. layer5 | Filter size: 5 | # filters: 50 | Stride: 1 | Activation: ReLU |
| Dense layer | Units: 1024 | | | Activation: ReLU |
| Output | Units: 1 | | | Activation: Linear |

TABLE II
NETWORK SPECIFICATIONS FOR BI-GRU [10]

| Layer | Network specifications | | | |
|---|---|---|---|---|
| Input | $N$-length sequence | | | |
| Conv. layer | Filter size: 4 | # filters: 16 | Stride: 1 | Activation: ReLU |
| Bi-GRU layer1 | Size: 64 | Merge: concat | | Activation: ReLU |
| Bi-GRU layer2 | Size: 128 | Merge: concat | | Activation: ReLU |
| Dense layer | Units: 128 | | | Activation: ReLU |
| Output | Units: 1 | | | Activation: Linear |

Note that convolutional layers are used in both seq2point architectures, with all layers activated by ReLU functions except the output. As the network layers go deeper, larger number of filters in convolutional layers in S2p and more units in Bi-GRU layers in Bi-GRU help to enrich features.

*C. Self-supervised Learning*

SSL performs promisingly in computer vision applications without labeled data [32], where visual features are learned from large-scale unlabeled data to avoid the cost of data annotations. In SSL, a pretext task should be pre-designed for the network with a loss function to learn features from unlabeled data set followed by supervised downstream tasks for enriching features. The general pipeline of SSL is shown in Fig. 3.

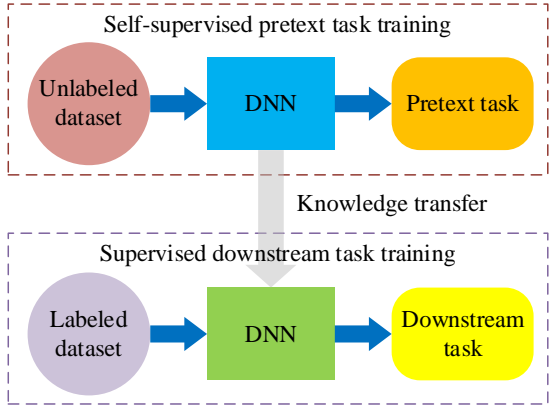

**Fig. 3.** SSL framework.

In Fig. 3, by solving the self-supervised pretext task, the network is pre-trained on unlabeled data set for mapping the relationship between the input and output in the pretext task. Then, benefiting from the downstream task, the pre-defined network parameters are fine-tuned based on the labeled data set. Thus, knowledge is transferred between tasks in SSL procedures.

### III. SELF-SUPERVISED LEARNING FOR NILM

Motivated by the drawbacks of existing DNN-based NILM methods requiring labeling for the target data set, in this paper, SSL is applied to various neural networks for load disaggregation. The scheme of SSL application in NILM is shown in Fig. 4, taking the S2p NILM method as an example. It can be observed that the proposed load disaggregation approach is composed of three stages: self-supervised pretext task training, supervised downstream task training, and load disaggregation as testing. For given aggregate power measurements, the S2p network can be constructed for mapping each $W$-length sliding window as a sequence $\mathbf{y}_{t:t+W-1}^{pre}$ to its midpoint $y_\tau^{pre}$ through pre-training. Note that self-supervised pre-training is treated as initialization for the general neural network $f$ for each appliance category, contributing to performance improvement and overfitting mitigation [33]. Then, for each appliance $m \in \mathcal{M}$, a load disaggregation network $f^m$ can be built by fine-tuning the pre-trained network $f$. That is, the network parameters are fine-tuned by representing each aggregate power sequence $\mathbf{y}_{t:t+W-1}^{fine}$ by the power consumed by such appliance at $x_\tau^m$. Finally, window-by-window aggregate power readings $\mathbf{y}_{t:t+W-1}^{test}$ for the target set are imported into $f^m$. Thus, the power consumed by each appliance is estimated sample-by-sample, corresponding to the midpoint $\hat{x}_\tau^m$ of each sequence for S2p. In the following part of this section, we clarify both self-supervised pretext task training and supervised downstream task training proposed for load disaggregation, while testing procedure is executed as in [2].

*A. Self-supervised Pretext Task Training*

Firstly, self-supervised pretext task training is set for extracting features from the unlabeled aggregate data from the target set. For matching the pre-training task with the network architectures of S2p and Bi-GRU, a non-linear loss function is defined as:

$$L_{SSL} = \min_{\theta_{SSL}} \frac{1}{l-W+1} \sum_{t=1}^{l-W+1} loss(\mathbf{y}_{t:t+W-1}^{pre}, y_0^{pre}), \quad (4)$$

where $\theta_{SSL}$ is a set of network parameters, and $y_0^{pre}$ is its midpoint for S2p or the ending point for Bi-GRU. Eq. (4) maps each aggregate power sequence referring to a sliding window of the mains to its midpoint or ending point via a non-linear regression.

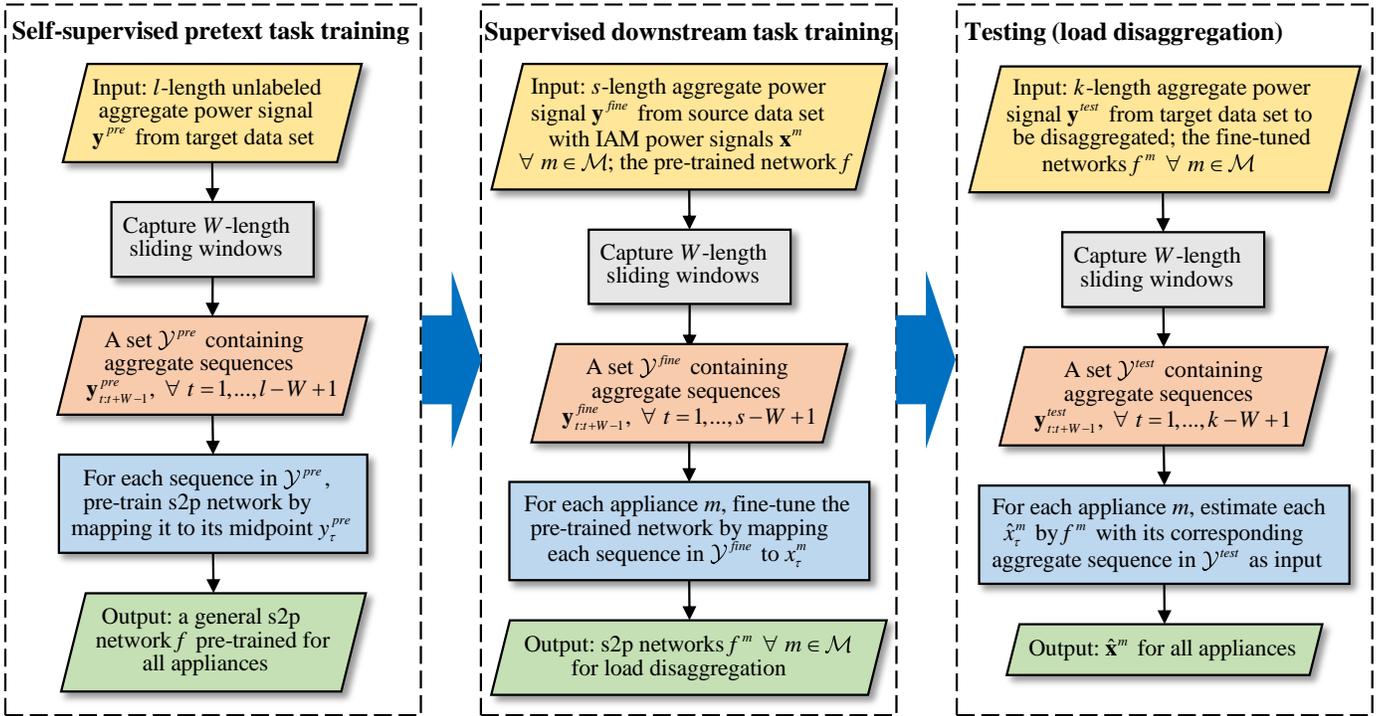

**Fig. 4.** Flow chart of the proposed SSL in load disaggregation, perfomed by S2p.

It is known that sudden changes within an aggregate power sequence correspond to appliance state transitions and switching ON/OFF events. Thus, the intuition behind setting such pretext task is to enhance feature representation, mainly referring to power changes [9]. Therefore, the network parameters are initialized via iterative updating in the self-supervised pre-training stage, with each sequence as input. The schematic diagram of the pretext task proposed for load disaggregation is shown in Fig. 5.

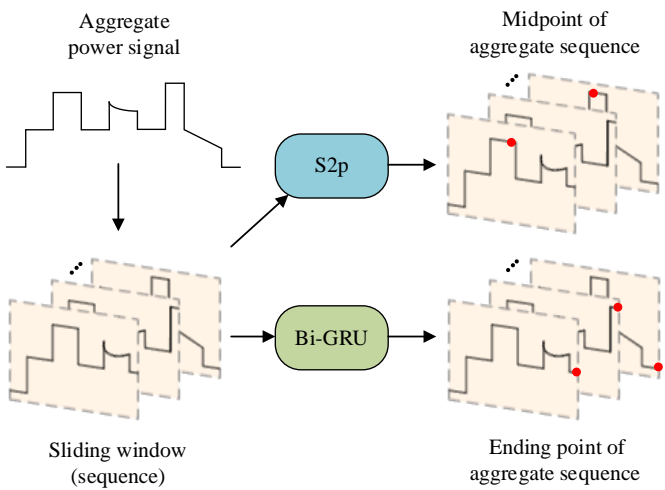

**Fig. 5.** Self-supervised pretext task layout.

As shown in Fig. 5, aggregate measurement sequences as sliding windows are iteratively imported into the networks, while the network outputs are their midpoints or ending points.

### B. Supervised Downstream Task Training

For transferring the features extracted from the 'unseen' target data set without labels to load disaggregation task, the network is updated iteratively. That is, the network parameters are further fine-tuned on the labeled data from the source set, as a supervised downstream task training. As described in Section II, such network is fine-tuned with the mains power in each sequence window as input and the appliance-level power estimation at its midpoint or ending point as output. Thus, the downstream task training for a typical appliance $m$ is illustrated in Fig. 6.

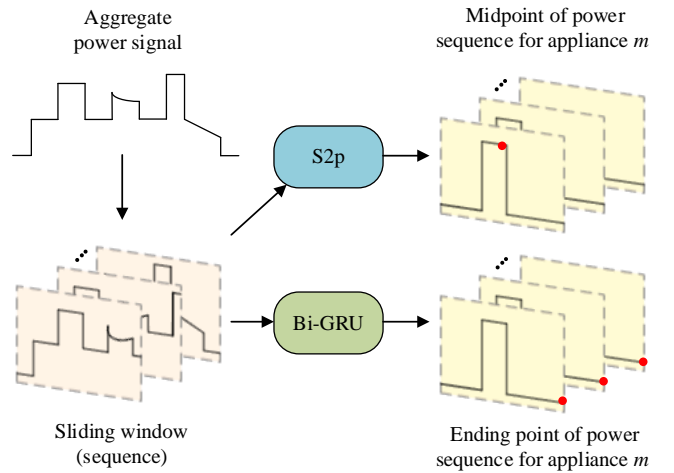

**Fig. 6.** Supervised downstream task layout.

Note that parameter fine-tuning can be carried out on either partial network layers or all layers, defined as partial fine-tuning and full fine-tuning. For performing partial fine-tuning on both S2p and Bi-GRU architectures, only their dense layers and output layers are fine-tuned, while the other layers are frozen. In DNN, general low-level features can be captured in shallow layers, while in deep layers, high-level features related to the task tend to be extracted [33]. Therefore, partial fine-tuning is usually applied to the tasks with knowledge transfer for sharing common features and thus guaranteeing efficiency. However, it is claimed in [34] that fine-tuning all network parameters after self-supervised pretext outperforms partial fine-tuning in some tasks. As a self-supervised pretext task and a supervised downstream task differ in inputs or outputs, the features extracted through the pretext task may not be suitable for the downstream task.

*C. Data Utilization in SSL against FSL & ZSL*

For clearly distinguishing data utilization in load disaggregation based on SSL, FSL, and ZSL, a comparison is demonstrated in Fig. 7.

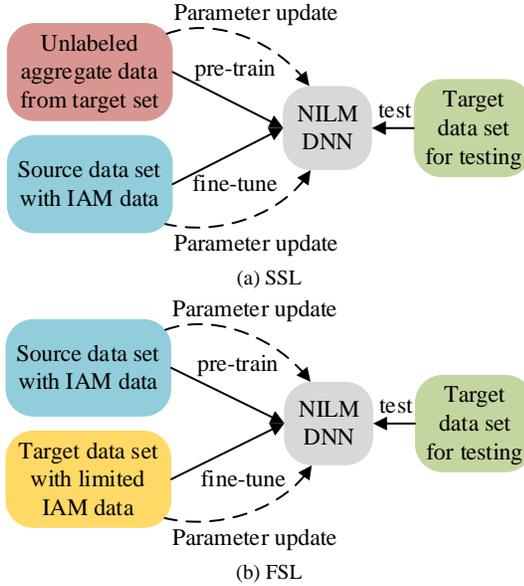

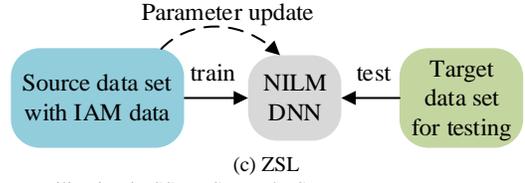

**Fig. 7.** Data utilization in SSL, FSL, and ZSL.

In an FSL-based NILM method, the network is pre-trained on the aggregate with sub-metering data in the source set and then fine-tuned on labeled data from the target set [30]. Namely, unlike in SSL, limited sub-metering data is required in FSL. However, in a ZSL-based NILM method, the network trained on the source data set is directly tested on the target data set, requiring no sub-metering measurements from the target set as in SSL.

## IV. EXPERIMENTAL SETUP

Based on S2p and Bi-GRU, SSL is benchmarked with ZSL in solving load disaggregation problems, as they neither require labeled data from the target set. Experiments are carried out on the US REDD data set [35] and two data sets from the UK, as UK-DALE [36] and REFIT [37].

*A. Data Selection*

The REDD data set [35] contains low-rate power readings collected from six houses in the United States for several weeks in 2011, sampled at 1 second. The UK-DALE data set [36], collected from five households in the UK from 2012 to 2017, includes the power consumption data sampled at 1 second. While in the REFIT data set [37], data is collected once per 8 seconds from 20 UK houses from 2013 to 2015. Comparing to the other two data sets, REFIT data set is noisier due to more unknown loads. Note that all data is down-sampled to 1 minute as low-rate measurements. Since the REFIT data set is the largest [2], containing more houses with more appliances, it is chosen for fine-tuning in SSL and training in ZSL. The experimental data selected for pretext task training, downstream task training, and testing in SSL is clarified in

TABLE III
DATA SELECTION FOR TESTING CASES

| Case | Pretext task training data set for SSL | | Samples ($\times 10^6$) | Downstream task training data set for SSL | Training data set for ZSL | Testing data set for both SSL and ZSL | | Samples ($\times 10^6$) |
|---|---|---|---|---|---|---|---|---|
| | House | Time period | | | | House | Time period | |
| 1 | REFIT | 2 | 2013-09-17 to 2015-02-06 | 0.54 | REFIT houses | Fridge: 5, 9, 12, 15<br>Microwave: 4, 10, 12, 17<br>Dishwasher: 5, 7, 9, 18<br>Waching machine: 5, 7, 10, 18<br>Kettle: 5, 7, 13, 20 | REFIT | 2 | 2015-02-06 to 2015-05-28 | 0.13 |
| 2 | | 8 | 2013-11-01 to 2015-05-10 | 0.72 | | | | | |
| | | 16 | 2014-01-10 to 2015-07-08 | 0.65 | | | | | |
| | | 19 | 2014-03-06 to 2015-06-20 | 0.63 | | | | | |
| 3 | REDD | 1 | 2011-04-18 to 2011-05-11 | 0.02 | | | REDD | 1 | 2011-05-11 to 2011-05-24 | 0.01 |
| 4 | | 2 | 2011-04-18 to 2011-05-22 | 0.02 | | | | | |
| | | 3 | 2011-04-16 to 2011-05-31 | 0.02 | | | | | |
| | | 4 | 2011-04-17 to 2011-06-04 | 0.03 | | | | | |
| | | 5 | 2011-04-18 to 2011-06-01 | 0.01 | | | | | |
| | | 6 | 2011-05-21 to 2011-06-14 | 0.02 | | | | | |
| 5 | UK-DALE | 1 | 2012-12-14 to 2017-04-26 | 2.24 | | | UK-DALE | 2 | 2013-05-20 to 2013-10-10 | 0.17 |
| 6 | | 1 | 2012-12-14 to 2016-06-02 | 1.77 | | | | | |
| | | 3 | 2013-02-27 to 2013-04-08 | 0.05 | | | | | |
| | | 4 | 2013-03-09 to 2013-10-01 | 0.22 | | | | | |
| | | 5 | 2014-06-29 to 2014-11-13 | 0.19 | | | | | |

Table III, as well as that for both training and testing in ZSL.

For exploring the performance boundary of SSL, six experimental cases are settled. In both Case 1 and Case 2, pre-training, fine-tuning, and testing in SSL are carried out among houses from the same REFIT data set. Case 1 and Case 2 share the same data from the same house for testing, while they differ in houses for building the pre-training data set. Data for single House 2 is used for both pre-training and testing in Case 1, while more houses other than House 2 are used for pre-training in Case 2. Similar settings are employed for Case 3 and Case 4, however, the data for both pre-training and testing is from REDD houses. For further investigating the impact of pre-training house selection on load disaggregation, we set the size of pre-training data from various houses in Case 6 equivalent to that from single House 1 in Case 5. In terms of ZSL benchmarks, the data sets for training and testing are the same as those for fine-tuning and testing for SSL, respectively. In consideration of the appliance types across data sets, five appliances with the most frequent usage are chosen to be disaggregated, including fridge, microwave, dishwasher, washing machine, and kettle. For each appliance, the data for fine-tuning its network comes from houses heuristically picked based on [28].

*B. Evaluation Metrics*

Since various metrics have been applied to evaluate NILM performance in the previous works, three widely-used metrics are selected in this paper. At first, mean absolute error ($MAE$) [4] is applied to quantify the absolute error between the predicted signal and the ground truth on average. Namely, $MAE$ is formulated as:

$$MAE = \sum_{t=1}^{T} |\hat{x}_t - x_t| / T, \qquad (5)$$

where the variable $\hat{x}_t$ denotes the power prediction at time index $t \in [1, T]$ and $x_t$ is the simultaneous actual power measurement. Thus, $T$ is the total number of samples in the signal.

The second metric is normalized signal aggregate error [28], defined as:

$$SAE = |\hat{r} - r| / r, \qquad (6)$$

where $r$ and $\hat{r}$ refer to the actual total energy consumption and its prediction, respectively. $SAE$ measures the relative error of the total energy.

Furthermore, metric energy per day ($EpD$) [2] is adopted which offers the average error of daily appliance-level energy consumption prediction. $EpD$ is defined as:

$$EpD = \sum_{n=1}^{D} |\hat{e}_n - e_n| / D, \qquad (7)$$

where $D$ denotes the total number of days, $e_n$ is the actual daily energy consumption, and $\hat{e}_n$ refers to the prediction. Not like $MAE$, where errors are calculated sample-by-sample, corresponding to power, $SAE$ and $EpD$ are calculated for energy consumption. However, $EpD$ reflects daily errors while $SAE$ is normalized for the entire period.

*C. Experimental Settings*

Our program is implemented in Python using TensorFlow. The networks were trained on machines with NVIDIA GeForce RTX 2080 Ti. Both networks are trained by ADAM optimizer algorithm, with an early-stopping mechanism to prevent overfitting as in [2], where the optimal network is picked with a validation loss lower than those in the following 6 epochs. The batch size is set to 512 for REDD data set, as in Case 3 and Case 4, while in other cases, it is set to 1024 for REFIT and UK-DALE data sets to accelerate data processing. Meanwhile, we set the learning rate to 0.001 as in [2]. The data normalization used in our experiments is the same as in [9]. In Bi-GRU, dropout rate is set to 0.5 and the window lengths are heuristically characterized as 10 for washing machines and 5 for other appliances, due to the lower sampling rate in our experiments. For the same reason, the general window length is set to 79 in S2p.

V. EXPERIMENTAL RESULT

In this section, the overall experimental results are initially demonstrated, as for all cases, SSL is evaluated against ZSL on both S2p and Bi-GRU networks in three metrics for each appliance. Then the total energy consumption per appliance estimated in all experiments is illustrated, followed by a series of examples of disaggregated power signals. Finally, comparisons between the training time for SSL on both networks and that for ZSL are shown. For simplification, SSL with partial fine-tuning and that with full fine-tuning are abbreviated as PSSL and FSSL, respectively. The abbreviations of such appliances are: F for fridge; M for microwave, DW for dishwasher, WM for washing machine, and K for kettle.

*A. NILM Performance Metrics Comparison*

The load disaggregation performance of ZSL, PSSL, and FSSL on both S2p and Bi-GRU networks for all cases is demonstrated in Table IV. For a clear expression, hyphens are added for combining disaggregation networks and the approaches they are trained.

*1) Within the same data set:* Since only REFIT data set is used in both Case 1 and Case 2, their results demonstrate load disaggregation performance of SSL and ZSL on both networks across houses from the same data set. In Case 1, S2p-FSSL and S2p-PSSL generally outperform the S2p-ZSL in various metrics, except for WM. It is mainly caused by the well-estimated power signal for WM close to the ground truth, especially for the operational states. In terms of K, although S2p-PSSL gains the best $MAE$, it suffers from over-estimation caused by unknown loads with similar power ranges. Bi-GRU-FSSL outperforms Bi-GRU-ZSL in overall $MAE$, although the later one achieves lower $MAE$ for most appliances. The main contributor for disaggregation performance by Bi-GRU-FSSL is WM, with less predicted stand-by power samples. Since PSSL performs the best in overall $SAE$ against others on both S2p and Bi-GRU networks, its appliance-level energy consumption estimation is generally the best in this case.

TABLE IV
THE EXPERIMENTAL RESULTS. BOLD NUMBERS REFER TO THE BEST RESULTS IN A ROW. THE RESULTS IN BRACKETS ARE FOR PSSL

| Case No. | App. | S2p-ZSL MAE (W) | SAE | EpD (kWh) | S2p-FSSL (S2p-PSSL) MAE (W) | SAE | EpD (kWh) | Bi-GRU-ZSL MAE (W) | SAE | EpD (kWh) | Bi-GRU-FSSL (Bi-GRU-PSSL) MAE (W) | SAE | EpD (kWh) |
|---|---|---|---|---|---|---|---|---|---|---|---|---|---|
| 1 | F | 28.10 | 0.06 | **0.12** | **26.69** (32.55) | 0.12 (0.39) | 0.16 (0.33) | 32.10 | 0.04 | 0.24 | 31.54 (32.29) | **0.01** (0.09) | 0.23 (0.28) |
|   | M | 11.94 | 1.66 | 0.16 | 11.27 (8.79) | 1.51 (0.66) | 0.15 (**0.08**) | **8.49** | **0.61** | 0.09 | 11.02 (8.88) | 1.30 (0.70) | 0.13 (0.09) |
|   | DW | 54.22 | 0.39 | 0.63 | **39.57** (43.64) | 0.13 (**0.03**) | 0.38 (**0.34**) | 87.83 | 0.59 | 1.22 | 95.15 (105.56) | 0.67 (0.70) | 1.19 (1.19) |
|   | WM | **35.06** | **0.13** | **0.40** | 45.64 (39.55) | 0.83 (0.41) | 0.52 (0.43) | 47.48 | 0.80 | 0.63 | 35.86 (41.69) | 0.16 (0.45) | 0.41 (0.46) |
|   | K | 25.87 | 0.59 | 0.30 | 18.55 (**18.00**) | 0.26 (0.16) | 0.16 (0.15) | 18.30 | 0.23 | 0.13 | 18.36 (24.04) | 0.20 (**0.07**) | 0.13 (**0.12**) |
|   | mean | 31.04 | 0.57 | 0.32 | **28.34** (28.51) | 0.57 (**0.33**) | **0.27** (**0.27**) | 38.84 | 0.45 | 0.46 | 38.39 (42.49) | 0.47 (0.40) | 0.42 (0.42) |
| 2 | F | 28.10 | 0.06 | **0.12** | 27.62 (28.73) | **0.01** (0.13) | 0.14 (0.19) | 32.10 | 0.04 | 0.24 | 31.19 (32.33) | **0.01** (0.14) | 0.24 (0.27) |
|   | M | 11.94 | 1.66 | 0.16 | 7.25 (9.46) | 0.25 (0.84) | 0.07 (0.09) | 8.49 | 0.61 | 0.09 | **5.57** (10.91) | **0.24** (1.24) | **0.05** (0.13) |
|   | DW | 54.22 | 0.39 | 0.63 | **45.48** (49.45) | **0.14** (0.26) | 0.39 (0.46) | 87.83 | 0.59 | 1.22 | 89.70 (104.85) | 0.60 (0.72) | 1.21 (1.21) |
|   | WM | 35.06 | 0.13 | 0.40 | 34.00 (**33.92**) | 0.18 (0.09) | 0.42 (**0.38**) | 47.48 | 0.80 | 0.63 | 34.89 (43.23) | **0.08** (0.54) | 0.43 (0.49) |
|   | K | 25.87 | 0.59 | 0.30 | 19.08 (20.23) | 0.22 (0.19) | 0.14 (0.15) | **18.30** | 0.23 | **0.13** | 19.87 (28.75) | 0.20 (**0.03**) | 0.13 (0.14) |
|   | mean | 31.04 | 0.57 | 0.32 | **26.69** (28.36) | **0.16** (0.30) | 0.23 (0.25) | 38.84 | 0.45 | 0.46 | 36.24 (44.01) | 0.23 (0.53) | 0.41 (0.45) |
| 3 | F | 41.74 | 0.15 | 0.21 | 43.62 (**36.70**) | 0.23 (**0.09**) | 0.29 (**0.16**) | 46.30 | 0.48 | 0.57 | 45.59 (50.51) | 0.48 (0.54) | 0.57 (0.64) |
|   | M | 18.65 | **0.26** | 0.13 | 18.78 (17.79) | **0.26** (0.55) | **0.12** (0.24) | **16.88** | 0.58 | 0.24 | 17.32 (18.86) | 0.65 (0.63) | 0.26 (0.26) |
|   | DW | 57.08 | 0.18 | 0.20 | 51.13 (61.05) | **0.02** (0.25) | 0.18 (0.25) | **40.04** | 0.51 | 0.24 | 42.62 (66.92) | 0.41 (0.54) | 0.27 (0.48) |
|   | WM | 30.49 | 0.76 | 0.65 | **30.11** (37.00) | 0.63 (0.55) | 0.61 (0.73) | 38.19 | 0.45 | 0.68 | 32.58 (40.25) | 0.60 (**0.39**) | 0.62 (0.70) |
|   | mean | 36.99 | 0.34 | **0.30** | 35.91 (38.14) | **0.29** (0.36) | 0.30 (0.35) | 35.35 | 0.51 | 0.43 | **34.53** (44.14) | 0.54 (0.53) | 0.43 (0.52) |
| 4 | F | 41.74 | 0.15 | **0.21** | 38.74 (**35.46**) | 0.32 (**0.13**) | 0.39 (**0.21**) | 46.30 | 0.48 | 0.57 | 47.32 (50.91) | 0.54 (0.57) | 0.64 (0.68) |
|   | M | 18.65 | **0.26** | **0.13** | 18.51 (18.10) | 0.42 (0.47) | 0.18 (0.20) | **16.88** | 0.58 | 0.24 | 17.32 (19.27) | 0.45 (0.63) | 0.19 (0.26) |
|   | DW | 57.08 | 0.18 | 0.20 | 40.82 (56.97) | 0.43 (**0.07**) | 0.17 (**0.14**) | **40.04** | 0.51 | 0.24 | 40.11 (64.71) | 0.52 (0.43) | 0.26 (0.42) |
|   | WM | 30.49 | 0.76 | 0.65 | **28.14** (36.18) | 0.71 (0.62) | **0.60** (0.75) | 38.19 | 0.45 | 0.68 | 33.16 (37.20) | 0.46 (**0.34**) | 0.61 (0.65) |
|   | mean | 36.99 | 0.34 | **0.30** | 31.55 (36.68) | 0.47 (**0.32**) | 0.34 (0.33) | 35.35 | 0.51 | 0.43 | 34.44 (43.02) | 0.49 (0.49) | 0.43 (0.50) |
| 5 | F | 19.96 | 0.23 | 0.25 | 21.32 (**18.40**) | 0.26 (**0.15**) | 0.29 (**0.18**) | 25.45 | 0.24 | 0.27 | 25.97 (28.21) | 0.27 (0.25) | 0.29 (0.28) |
|   | M | 12.78 | 0.68 | 0.11 | 11.34 (11.43) | 0.54 (**0.14**) | 0.09 (**0.06**) | 8.86 | 0.17 | **0.06** | **7.71** (8.88) | 0.40 (0.38) | 0.08 (0.08) |
|   | DW | 36.11 | 0.34 | 0.35 | 18.54 (26.66) | 0.21 (0.29) | 0.25 (0.33) | **14.65** | 0.15 | 0.19 | 18.48 (28.38) | **0.02** (0.09) | 0.14 (0.23) |
|   | WM | 17.21 | 0.15 | **0.22** | 15.84 (17.03) | 0.11 (0.20) | 0.24 (0.26) | 17.60 | 0.26 | 0.29 | 15.75 (**15.68**) | **0.04** (0.05) | 0.25 (0.27) |
|   | K | 17.57 | **0.08** | **0.10** | **13.37** (13.84) | 0.13 (0.25) | **0.10** (0.18) | 17.67 | 0.50 | 0.36 | 18.42 (20.52) | 0.53 (0.54) | 0.37 (0.38) |
|   | mean | 20.73 | 0.30 | 0.21 | **16.08** (17.47) | 0.25 (**0.21**) | 0.19 (0.20) | 16.85 | 0.26 | 0.23 | 17.27 (20.33) | 0.25 (0.26) | 0.23 (0.25) |
| 6 | F | 19.96 | 0.23 | 0.25 | 21.19 (**18.33**) | 0.25 (**0.11**) | 0.27 (**0.15**) | 25.45 | 0.24 | 0.27 | 24.31 (27.29) | 0.18 (0.22) | 0.20 (0.24) |
|   | M | 12.78 | 0.68 | 0.11 | **6.93** (11.15) | 0.28 (**0.10**) | **0.06** (**0.06**) | 8.86 | 0.17 | **0.06** | 7.16 (8.69) | 0.50 (0.41) | 0.09 (0.08) |
|   | DW | 36.11 | 0.34 | 0.35 | 25.11 (36.63) | 0.04 (0.02) | 0.17 (0.24) | **14.65** | 0.15 | 0.19 | 21.02 (24.19) | **0.01** (0.17) | **0.15** (0.25) |
|   | WM | 17.21 | 0.15 | **0.22** | 15.49 (32.73) | 0.16 (1.59) | 0.26 (0.49) | 17.60 | 0.26 | 0.29 | 18.17 (15.90) | 0.30 (**0.08**) | 0.29 (0.27) |
|   | K | 17.57 | **0.08** | **0.10** | 13.33 (17.86) | 0.12 (0.19) | **0.10** (0.15) | 17.67 | 0.50 | 0.36 | 18.14 (25.92) | 0.47 (0.43) | 0.33 (0.31) |
|   | mean | 20.73 | 0.30 | 0.21 | **16.41** (23.34) | **0.17** (0.40) | **0.17** (0.22) | 16.85 | 0.26 | 0.23 | 17.76 (20.40) | 0.29 (0.26) | 0.21 (0.23) |

Similar results can be observed in Case 2, where the data for pre-training is acquired from other more houses instead of the same single house for testing as in Case 1. In Case 2, S2p-FSSL performs the best in all overall metrics. However, better disaggregation results for M are achieved by Bi-GRU-FSSL due to less mis-identification. Therefore, in both Case 1 and Case 2, we can conclude that FSSL generally outperforms ZSL and PSSL in the load disaggregation tasks addressed on both S2p and Bi-GRU. In short, more pre-training data generally helps to improve load disaggregation performance.

*2) Across data sets:* In both Case 3 and Case 4, the data set for both training in ZSL and fine-tuning in SSL remains the same as in Case 1 and Case 2, while the data in other stages is switched to REDD data set. *EpD* of S2p-FSSL is lower for most appliances while comparable with S2p-ZSL on average. Although ZSL achieves lower *MAE* in Bi-GRU for most appliances, Bi-GRU-FSSL achieves lower *MAE* for WM, resulting in better overall *MAE* performance. The load disaggregation performance for F is improved by pre-training, with close performance in Case 3 and Case 4, which is similar to the results achieved for WM. Note that S2p-PSSL outperforms S2p-FSSL for F, as overfitting may occur as F is a simple two-state appliance. However, for more complex WM loads containing multiple operational states, fine-tuning all network parameters may contribute to feature learning and better disaggregation results. The worst performance for DW among all appliances is caused by the difference between the DW load characteristics in the REDD data set and those in the REFIT data set. Such distinction in load characteristics also affects the disaggregation results for M. *MAE* results for both M and DW are increased by applying SSL to Bi-GRU, while in Case 3, the superiority of S2p-FSSL in both *SAE* and *EpD* is caused by lower power estimation for mis-identified loads. It is noticeable that PSSL generally improves *SAE* performance of both NILM methods in Case 4, especially for DW and WM, where enhancement in power estimation leads to closer energy consumption estimation performance. Since FSSL generally outperforms others in Case 3 and Case 4, the results show the knowledge transferability between REDD data set and REFIT data set.

For the remaining experiments in Case 5 and Case 6, data for self-supervised pre-training in SSL and testing in both ZSL and SSL is selected from UK-DALE data set. From Table IV, applying SSL to S2p generally achieves the best disaggregation performance in both cases. However, improvement is tiny for Bi-GRU. Moreover, replacing the data collected from a single house by that from multiple houses while maintaining the data size seems not to have a certain impact on load disaggregation performance. E.g., close performance for fridges in Case 5 and Case 6 is due to the high similarity in power characteristics of their operational cycles. Significant performance improvement is achieved by SSL for M from Case 5 and Case 6, due to less mis-identification. Bi-GRU-FSSL outperforms Bi-GRU-ZSL for DW in energy estimation, however, Bi-GRU-ZSL achieves better results in sample-by-sample disaggregation results.

From the whole results across cases, benefited from self-supervised pre-training, FSSL generally outperforms ZSL and PSSL in both networks, especially in S2p. FSSL generally outperforms PSSL, where feature adaption to NILM task in PSSL is constrained by freezing partial network parameters. Furthermore, S2p generally performs better than Bi-GRU across metrics. With respect to transferability between data sets, the overall results show UK-DALE and REFIT data sets, both collected from the UK, have stronger transferable potential than the data sets from distinct regions. Moreover, the increase in pre-training data size leads to performance improvement for SSL. In terms of disaggregation performance for each appliance, the poor results for M are achieved in various cases, as little information can be captured due to its short duration and low usage frequency. For more complex WM loads with multiple operational states, their characteristics differ a lot across houses and data sets, resulting in under-estimation for power ranges while over-estimation for operation times.

### B. Energy Consumption Estimation

The power consumption estimation results for each appliance are demonstrated as in Fig. 8. In both Case 1 and Case 2, power consumption close to the ground truth is estimated by various methods for F, M, and K, with ON/OFF operational states. However, they differ in power consumption estimation for multi-state DW and WM, respectively, where Bi-GRU suffers from over-estimation. Around 8% and 16% samples in power signals for DW and WM in the fine-tuning sets are in stand-by states between 0~10Watt, collected from both REFIT House 5 and House 7. The stand-by power samples are estimated by Bi-GRU for DW and WM, while they cannot be observed in the testing set. Similar results can be observed in Case 3 and Case 4. Moreover, the generally worse performance of Bi-GRU than S2p in estimating power consumption for F in Case 3 and Case 4 is also caused by the distinction between the power characteristics of F in fine-tuning set and those for testing set. The worse performance of Bi-GRU as under-estimation for K in Case 5 and Case 6 can be explained by the same reason. Thus, we can conclude that Bi-GRU is more sensitive to the distinction between distribution in training and testing sets. Note that unknown loads consume 73% in Case 1 and Case 2, 65% in Case 3 and Case 4, and 56% in Case 5 and Case 6 of the total energy, thus over-estimation is more serious in Case 1 and Case 2 due to more mis-identification.

Among all six cases, S2p generally outperforms Bi-GRU in total energy estimation, in line with the results shown in Table IV. Besides, SSL helps to improve both networks in power

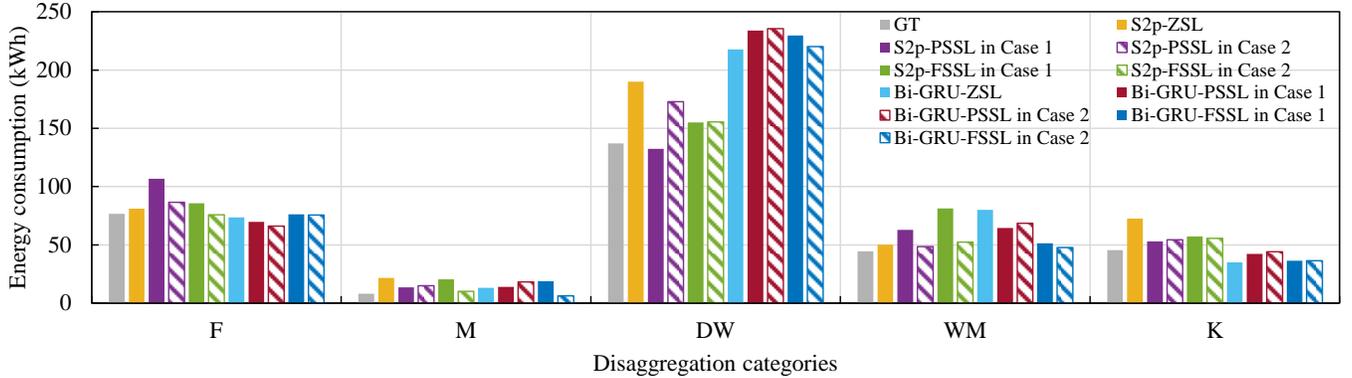
(a) Case 1 and Case 2

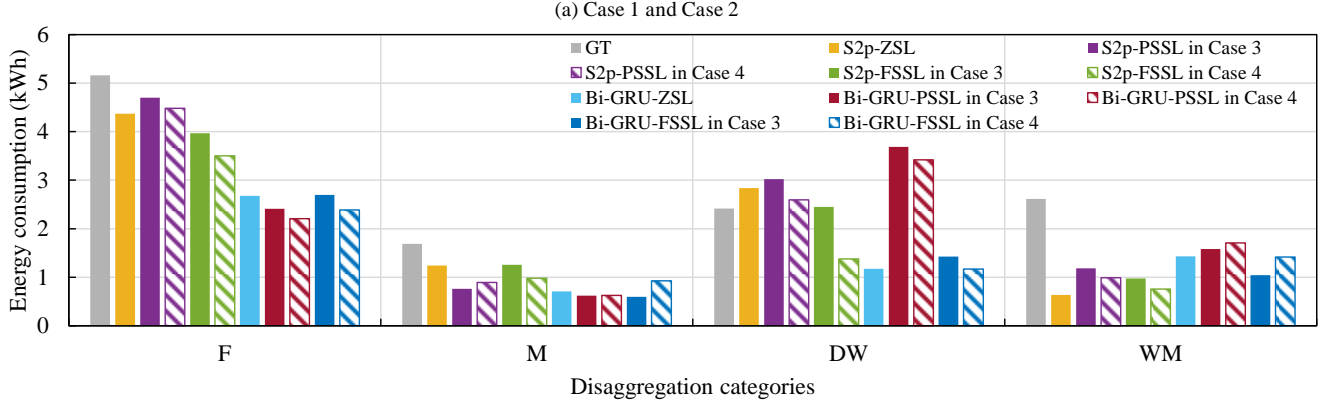
(b) Case 3 and Case 4

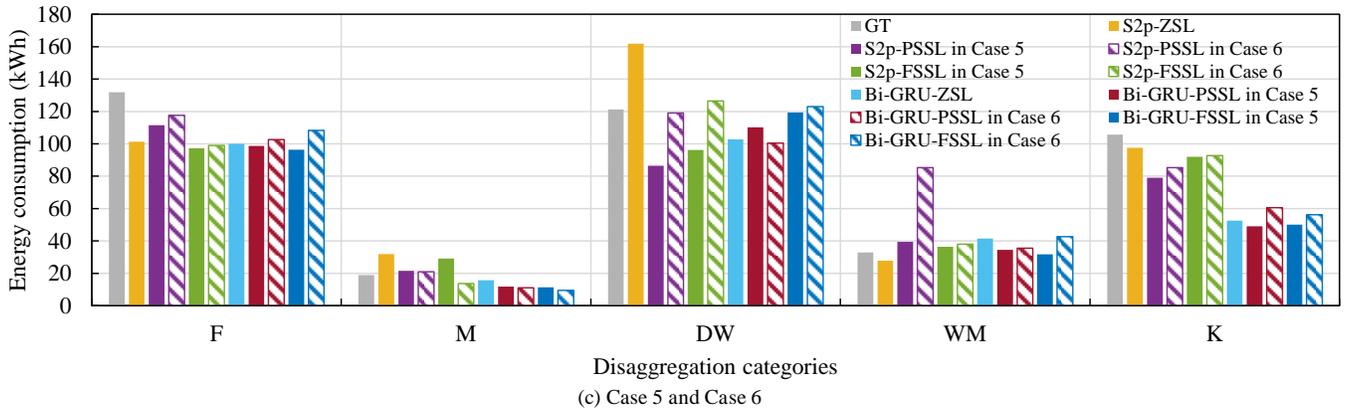

(c) Case 5 and Case 6

**Fig. 8.** Energy consumption estimation by ZSL, PSSL and FSSL with ground truth (GT).

consumption estimation performance, with slight superiority of S2p-FSSL comparing to S2p-PSSL.

### C. Load Power Disaggregation

Furthermore, Case 6 is selected as an example for demonstrating load disaggregation results. The estimation of power consumed by five target appliances in Case 6 is shown in Fig. 9. From Fig. 9, the operating periods for each target appliance are generally identified by these methods. Furthermore, the superiority of NILM methods based on S2p over those based on Bi-GRU can be observed, with disaggregated power signals closer to the ground truth for most appliances. Among these methods, under-estimation tends to be common for most appliances. Possible reasons can be inferred as the mapping construction is affected by short-lasting and less-frequent usage of M, and multi-state WM operating in various power ranges. An exception is dishwasher, which is over-estimated by both S2p-ZSL and S2p-PSSL, as DWs from REFIT houses operate in higher power ranges than those from UK-DALE houses.

### D. Training Time Analysis

In this sub-section, the total training time in both Case 4 and Case 5 is demonstrated in Fig. 10 as instances. In Case 4, the shortest training time is achieved by Bi-GRU-PSSL for all appliances, with two main reasons can be inferred. Firstly, since more layers are fine-tuned in FSSL than in PSSL, PSSL is generally carried out faster under the same conditions. Then, network parameters in Bi-GRU are fewer than those in S2p architecture. Therefore, more training time required by S2p-FSSL is explainable. Especially, due to fridges from pre-training set and fine-tuning set differ in operational patterns as power ranges and periodicity, more epochs are required for network convergence, thus lead to longer duration. Since M operation usually lasts within several minutes, its distinction on power profile across data sets and houses is trivial. Therefore, close training time is achieved for M. However, the amount of data for pre-training in Case 5 is above 20 times of that in Case 4, resulting in the superiority of ZSL on both models in training time per appliance. It is noteworthy that applying SSL to Bi-GRU requires longer training time for multi-state appliances such as WM, due to its greatest sliding window length among all appliances in Bi-GRU. In conclusion, the amount of data for pre-training should be carefully chosen as it balances NILM performance and efficiency.

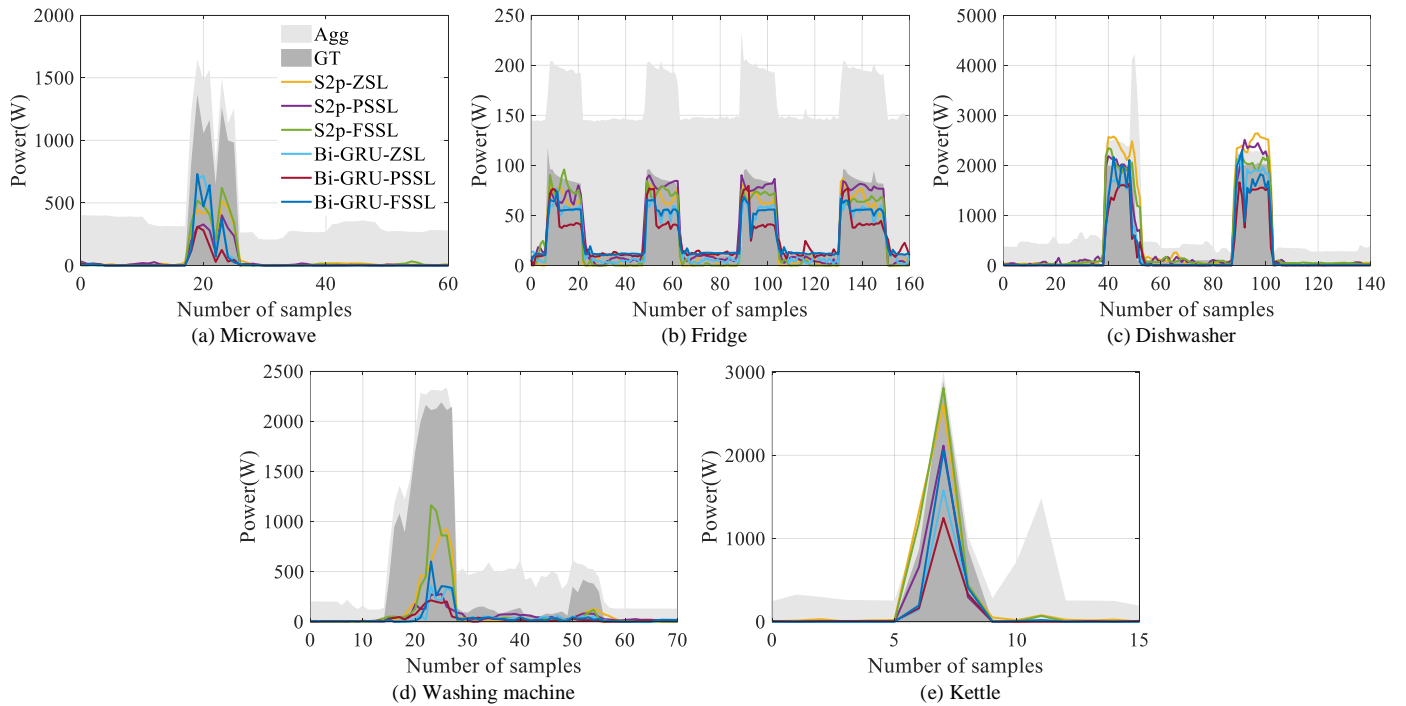

**Fig. 9.** Disaggregated power signals for UK-DALE House 2 in Case 6.

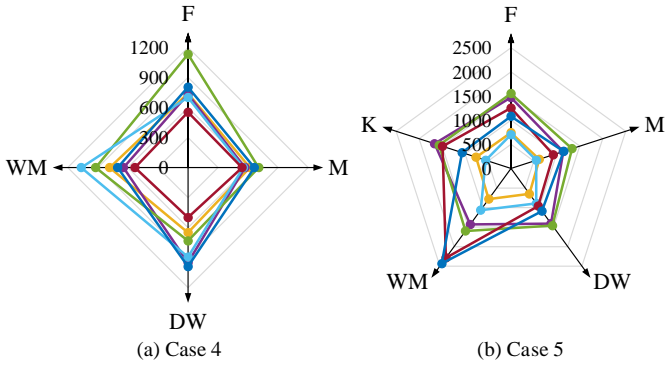

**Fig. 10.** The total training time (in second) for Case 4 and Case 5.

## VI. CONCLUSION AND FUTURE WORK

In this paper, SSL is applied to NILM solutions based on the state-of-the-art S2p and Bi-GRU. A pretext task is proposed to pre-train a general network to map aggregate power sequences to derived representatives, where labeled data for the target data set is not required. By performing a supervised downstream task based on the labeled data from source data sets, the pre-trained network is fine-tuned with feature transference and then applied to disaggregate loads for the target sites. Publicly-accessible REDD, UK-DALE, and REFIT data sets are utilized for validation, with multiple experimental cases designed to evaluate the performance of SSL using data within the same set and across data sets. For a comprehensive NILM performance demonstration, various metrics are employed, showing SSL (especially FSSL) generally outperforms ZSL scheme. Specifically, the features learned via SSL help to improve under-estimation, reduce mis-identification and mitigate the influence of wrong labels. Such conclusions are also supported by the disaggregated power signals for each appliance and their total power consumption estimation results. Thus, SSL contributes to network refinement and disaggregation performance improvement. Furthermore, FSSL performs better than PSSL in disaggregation accuracy, although takes longer execution time.

Future work includes investigation on the influence of activation balancing for compensating the skew of training data; analyzing the impact of sampling frequency of the data for fine-tuning in the source domain on the NILM performance; and applying autoencoder to carry out the pretext task.